\documentclass[aps,prl,showpacs,twocolumn,groupedaddress,amssymb]{revtex4}

\usepackage{graphicx}
\usepackage{dcolumn}
\usepackage{bm}
\usepackage{subfigure}

\begin{document}

\title{Kink-induced transport and segregation in oscillated granular layers}
\author{Sung Joon Moon}
\email[]{moon@chaos.utexas.edu}
\author{Daniel I. Goldman}
\author{J. B. Swift}
\email[]{swift@chaos.utexas.edu}
\author{Harry L. Swinney}
\affiliation{Center for Nonlinear Dynamics and Department of Physics,
          University of Texas, Austin, Texas 78712}
\date{\today}

\begin{abstract}

We use experiments and molecular dynamics simulations of vertically
oscillated granular layers to study horizontal particle segregation
induced by a kink (a boundary between domains oscillating out of phase).
Counter-rotating convection rolls carry the larger particles in
a bidisperse layer along the granular surface to a kink, where they
become trapped. The convection originates from avalanches that occur
inside the layer, along the interface between solidified and fluidized
grains. The position of a kink can be controlled by modulation of
the container frequency, making possible systematic harvesting of
the larger particles.

\end{abstract}


\maketitle
\nobreak

In 1831 Faraday observed convective motion of grains in heaps
in vertically oscillated granular layers~\cite{faraday}.
Later there were many studies of convection in oscillated
granular materials~\cite{kroll,ratkai,savage,laroche,pak,clement,gallas,
taguchi,lee,chicago,wildman}. These convection phenomena are driven by
interstitial air~\cite{laroche,pak} and shear due to
sidewalls~\cite{clement,lee,chicago}.
One consequence of the sidewall-driven convection is vertical
segregation of grains of different sizes, which is an example of
the Brazil-nut effect~\cite{brown39,williams76,moebius01}.

We consider here a different kind of convection, one that is driven
not by interstitial air or sidewalls but rather arises from the {\em
intrinsic} dynamics of the oscillated granular layers. This convection
is associated with kinks, which are boundaries separating oscillating
domains of opposite phase; when the layer on one side of a kink is
moving up, the layer on the other side of the kink is moving
down. Kinks can form in oscillating layers when the maximum container
acceleration $a_{max}$ is large enough ($a_{max} \geq~4.5g$, where $g$
is the gravitational acceleration) so that the granular layer hits the
container bottom only every other cycle~\cite{melo95,umbanhowar98}.
Kinks spontaneously form in a layer for $a_{max} > 7g$~\cite{moon02a}.

During each oscillation cycle, the granular material on one side of
a kink strikes the container bottom and solidifies,
while the fluidized grains on the other side
avalanche down along the front of the solidified grains.
We examine this avalanche process, which will be shown to be
responsible for convection associated with kinks.
In a bidisperse layer, such convective motion conveys larger grains
toward the kink.
We will show that controlled motion of a kink leads to horizontal
size segregation, in contrast to the vertical size segregation in
the Brazil-nut effect. We will first describe the laboratory
observations, and then show how the particle transport, trapping,
and horizontal segregation can be understood using molecular
dynamics simulations.

\begin{figure}[b!h]
\begin{center}
\includegraphics[width=.83\columnwidth]{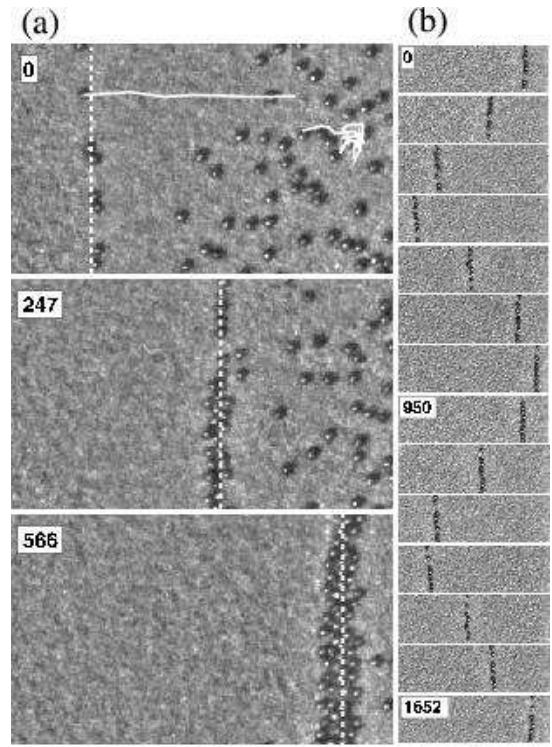}
\end{center}
\caption{\label{Transport} Segregation and controlled transport of
$650~\mu$m diameter black glass spheres by a kink in an experiment with a 10-particle deep oscillated layer of $165~\mu$m bronze spheres
at $a_{max}=5g$ and $f$ = 92 Hz. The motion of the kink is controlled
by modulation of the phase difference $\Delta\phi$ between the primary
oscillation signal and an added small subharmonic sinusoidal perturbation
(see text). (a) A kink (dashed line) sweeps across the layer after
a rapid change in $\Delta\phi$ of $2\pi/3$; glass spheres close to
the kink move toward it and remain trapped in the kink at the
surface of the layer, while far from the kink they diffuse; these two
types of trajectories are illustrated by the solid white lines.
The numbers in each frame denote plate oscillations after $\Delta\phi$
is changed. (b) Sinusoidal oscillation of a kink, with
$\Delta\phi(t)=f_{ms}/f_{mr} \sin(2\pi f_{mr} t)$, $f_{ms} = 0.17$ Hz
and $f_{mr} =0.1$ Hz. The large trapped particles follow the motion
of the kink.}
\end{figure}

\begin{figure}[t]
{\includegraphics[width=.8\columnwidth]{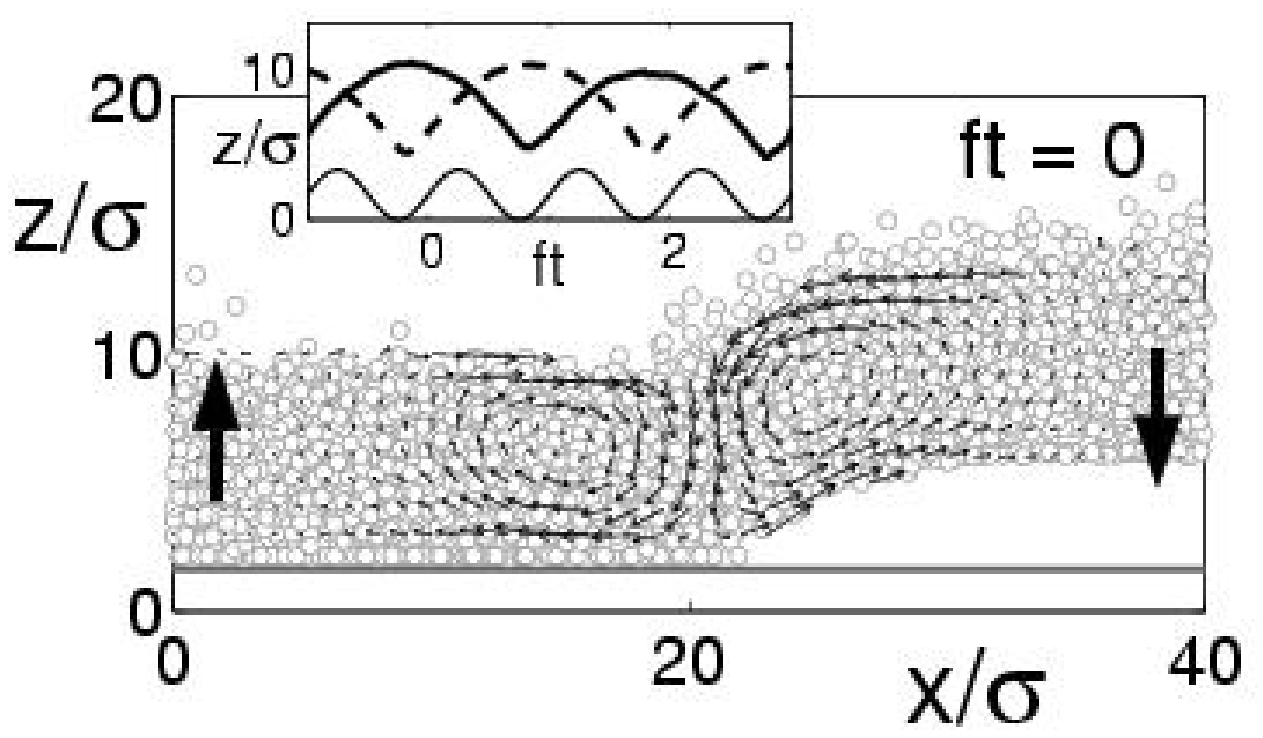}
\includegraphics[width=.49\columnwidth]{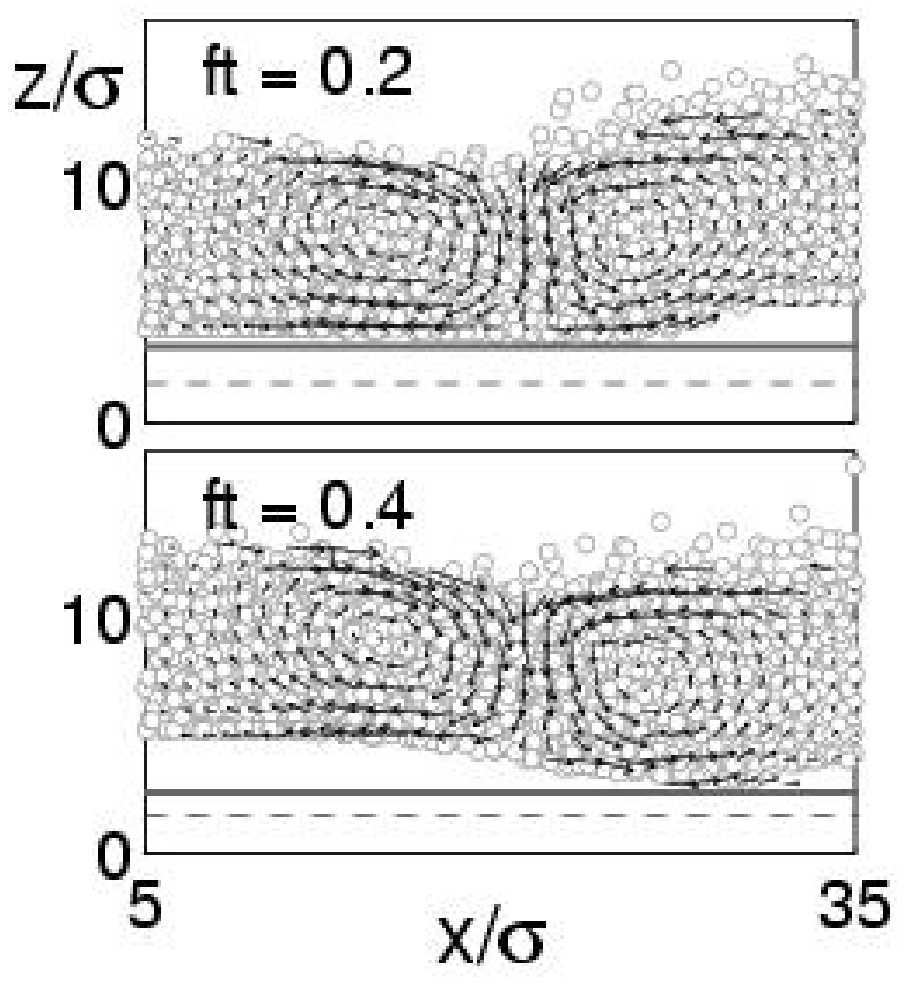}
\includegraphics[width=.49\columnwidth]{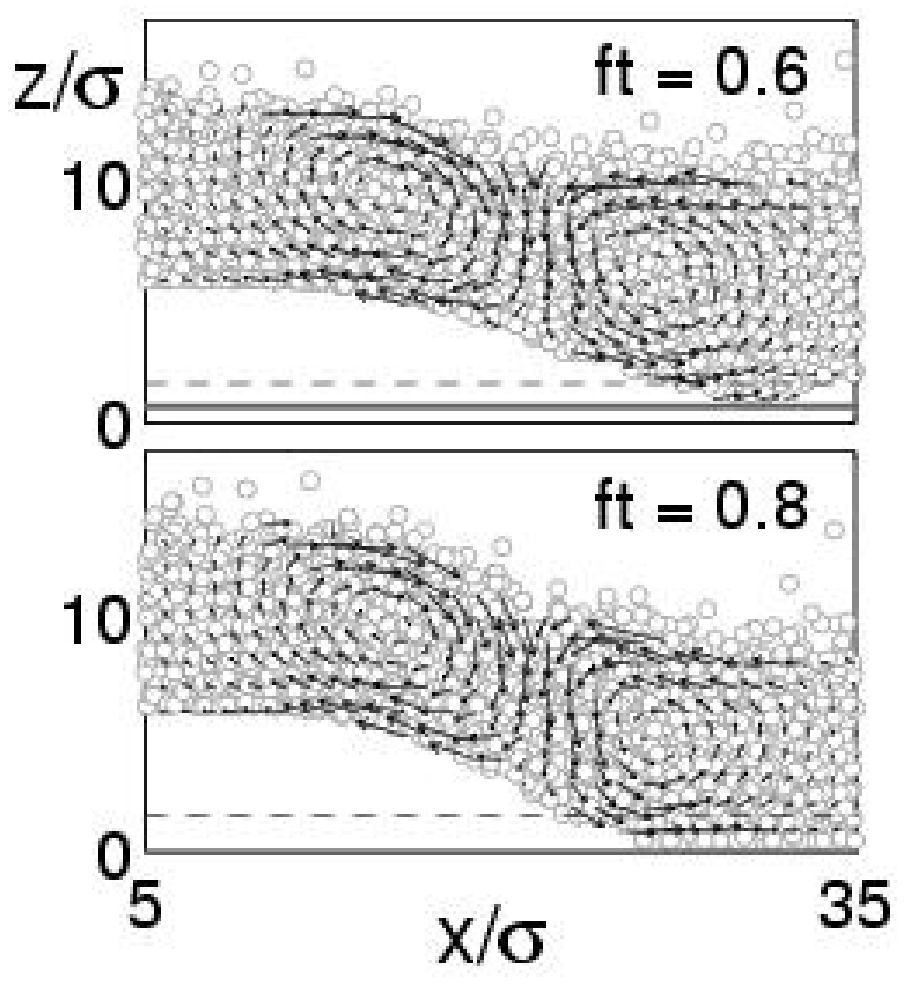}}
\caption{ \label{ShowConvection} Grains form a pair of convection rolls,
flowing downward at the kink, as illustrated by these projections of
a layer with a kink at different times $ft$ during a cycle ($ft=0$
when the container is at its equilibrium position, moving upward),
obtained from a simulation ($a_{max} = 5.2g$, $f = 69~{\rm Hz}$).
The vertical position of the center of mass
as a function of time of the left (right) side of the kink is shown
as dashed (solid) line in the inset of the top panel. The two bold
vertical arrows on either side of the kink indicate the directions of
motion of these two domains. The small arrows show the discrete grain
velocity ${\bf u}({\bf r}(t),t)$ averaged over the shorter horizontal
direction. Each circle represents a grain, and the container bottom is
indicated by horizontal gray lines. The region shown is far from
the rigid sidewalls in the longer horizontal direction.
}

\end{figure}

{\it Experimental observations: Trapping, transport, and segregation ---}
We find that larger grains in the vicinity of a kink in an oscillating
bidisperse layer move toward the kink and remain there, while the larger
grains far from a kink move only diffusively, as the solid white lines
in Fig.~\ref{Transport} (a) illustrate. Further, if the location of
a kink is controlled by modulating the container oscillation
frequency~\cite{aranson99}, the large grains follow the motion of
the kink (Fig.~\ref{Transport} (b)); this leads to a horizontal size
segregation. We consider a 10-particle deep layer of 165 $\mu$m bronze
spheres (mass density $8.3~\mbox{g}/\mbox{cm}^3$)
together with a few hundred of 650 $\mu$m glass spheres
($2.5~\mbox{g}/\mbox{cm}^3$), in an evacuated square container of
horizontal area $8.9 \times 8.9~\mbox{cm}^2$ (the pressure is 5 Pa).
The layer is subject to a vertical sinusoidal oscillation with an
amplitude $A$ and a frequency $f$; the oscillation is characterized by
two of the following three parameters: the maximum acceleration
$a_{max} = A(2\pi f)^2$, the maximum velocity $V_{max} = 2\pi Af$, and
the frequency $f$. In some experiments we add a perturbative subharmonic
forcing to control the position of a kink, where the plate position $z$
is given by $z(t)=A\sin(2\pi ft) +0.01A\sin(\pi ft+\Delta\phi(t))$.
This secondary forcing breaks the symmetry of the momentum transfer
across the kink, which makes the kink move~\cite{moon02a}. The relative
phase between the main oscillation and the subharmonic
perturbation controls the position of the kink within the container
(see Fig.~\ref{Transport}).

We consider the range $4.5g < a_{max} < 5.5g$, where the
layer remains flat except for the kinks; similar transport and
segregation occur for larger $a_{max}$ where wave patterns form,
but it is simpler to quantify the transport and segregation
processes for a non-patterned flat layer.

\begin{figure}[b]
\includegraphics[width=.91\columnwidth]{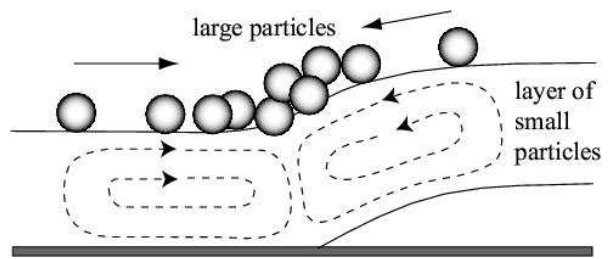}
\caption{
\label{TwoSizeLayer}
A schematic diagram showing trapping of larger particles at a kink
in a layer of smaller particles. The larger particles move toward
the kink due to the surface flow of the convective motion of the
smaller particles. The larger particles do not flow down in between the
convection rolls, but remain there, due to their size. Dashed arrows
represent the convection rolls formed by the smaller particles, and
solid arrows indicate the direction of the motion of large particles.}
\end{figure}

{\it Microscopic dynamics; kink-associated convection ---}
The experimental observations reveal that a mean flow toward a kink
exists in the vicinity of the kink. We use a previously validated
three-dimensional (3-d) molecular dynamics
simulation~\cite{bizon98,moon02a}, to understand the trapping
and transport produced by the microscopic dynamics.
The coefficient of restitution $e$ and the coefficient of friction
$\mu$ were set to 0.7 and 0.5 respectively, as in Ref.~\cite{bizon98}.
We simulate 8-particle deep monodisperse layers of spherical grains
of diameter 165 $\mu$m in a rectangular box of horizontal area
$160\sigma \times 10\sigma$, where $\sigma$ is the particle diameter.
Periodic boundary conditions are used in the shorter horizontal
direction ($y$-axis), and vertical rigid sidewalls are imposed in
the longer horizontal direction ($x$-axis). The rigid sidewalls and
the bottom plate are assumed to be made of the same material as grains.

We distinguish two types of velocities, the instantaneous grain
velocity ${\bf v}({\bf r}(t),t) = \lim_{\Delta t \rightarrow 0^+}
({\bf r}(t+\Delta t)-{\bf r}(t))/\Delta t$ and the {\em discrete}
grain velocity ${\bf u}({\bf r}(t),t) = ({\bf r}(t+2T)-{\bf
r}(t))/(2T)$, where ${\bf r}(t)$ is the position vector of a grain
at a given time $t$, and $2T = 2/f$ is the oscillation period of the
macroscopic state. The discrete velocity corresponds to the grain
velocity in the $f/2$-strobed frame. Following the trajectories of
individual grains, we observe that grains do not come back to the
same positions after $2T$, even though the layer returns to the
same macroscopic state. Instead, in the discrete velocity field,
grains form a pair of counter-rotating convection rolls that flow
downward at the kink (Fig.~\ref{ShowConvection}). The shape of the
rolls changes during a cycle, as does the shape of the layer, but
the sense of circulation remains the same. For the parameters of
Fig.~\ref{ShowConvection}, it takes about 100$T$ for grains in the
outer-most part of the rolls to complete a full circuit.

{\it Trapping and transport ---} The presence of the convection rolls
explains the motion of large grains toward a kink: Large grains far
from the kink rise to the surface as a consequence of the Brazil-nut
effect and those that enter the convecting region will move toward
the kink due to the surface flow of the convective motion
(Fig.~\ref{TwoSizeLayer}). When the large grain reaches the down-flow
channel region between the convection rolls, it will not flow down
because it is too large, and it becomes trapped there, as illustrated
in Fig.~\ref{Transport}. We have observed trapping for a wide range of
impurity particle diameters and densities relative to those of the
oscillated bronze spheres. We introduced hundreds of the same
type of impurity particles, including lead, stainless steel,
glass, and polystyrene spheres of diameter ratios 1.12 to 11.8,
density ratios 0.14 to 1.4, and mass ratios 0.4 to 2500.
Trapping occurred in all cases studied except for the diameter ratio
below 1.12, where the motion of the impurity grains was visually
indistinguishable from the motion of the bronze spheres.

\begin{figure}[t]
\includegraphics[width=.87\columnwidth]{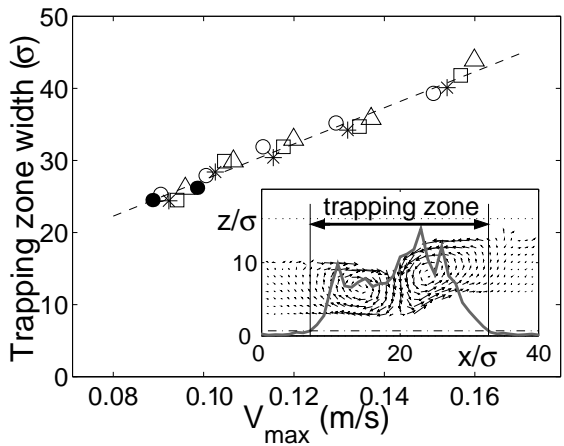}
\caption{ \label{VortexMeasure} The trapping zone width (defined in
the inset) for various $a_{max}$'s and $f$'s, obtained in simulations
of 8-particle deep monodisperse layers of 165 $\mu$m spheres.
The width is linearly proportional to $V_{max}$, as shown by
the dashed line (a least square fit). The symbols correspond to:
$\bullet$, $a_{max} = 4.9g$; $\circ$, $5g$; $\ast$, $5.1g$; $\square$,
$5.2g$; $\triangle$, $5.3g$. For each $a_{max}$, the data points
correspond to $f$ = 86 Hz, 78 Hz, 69 Hz, 60 Hz, 52 Hz,
respectively. Inset: The trapping zone is defined as the region where
the vertically integrated $|{\bf u}|^2$ (gray solid line, in an
arbitrary unit) is larger than some small value (dash-dotted line). }
\end{figure}

We refer to a region in which large grains move toward a kink and
become trapped as the ``trapping zone''. The width of the trapping
zone, defined in Fig.~\ref{VortexMeasure}, is found to be linearly
proportional to $V_{max}$ for the range of oscillation parameters in
our study (Fig.~\ref{VortexMeasure}).
There is a limit to the number of large grains that can be trapped;
for the range of $V_{max}$ studied in this paper,
the maximum number of trapped particles corresponds to the number of
large grains forming a monolayer above the trapping zone.
In the case of Fig.~\ref{Transport} (a), $V_{max}$ of the primary
oscillation is 0.087 m/s, and the trapping zone width from the last
frame ($t = 566T$) is seen to be about 5 glass particle diameters
or 20 bronze particle diameters, which is comparable to the simulation
result in Fig.~\ref{VortexMeasure}.

\begin{figure}[t]
{\sf \hskip .1\columnwidth (a) Lab. frame \hskip .22\columnwidth (b) Container frame}

\includegraphics[width=.493\columnwidth]{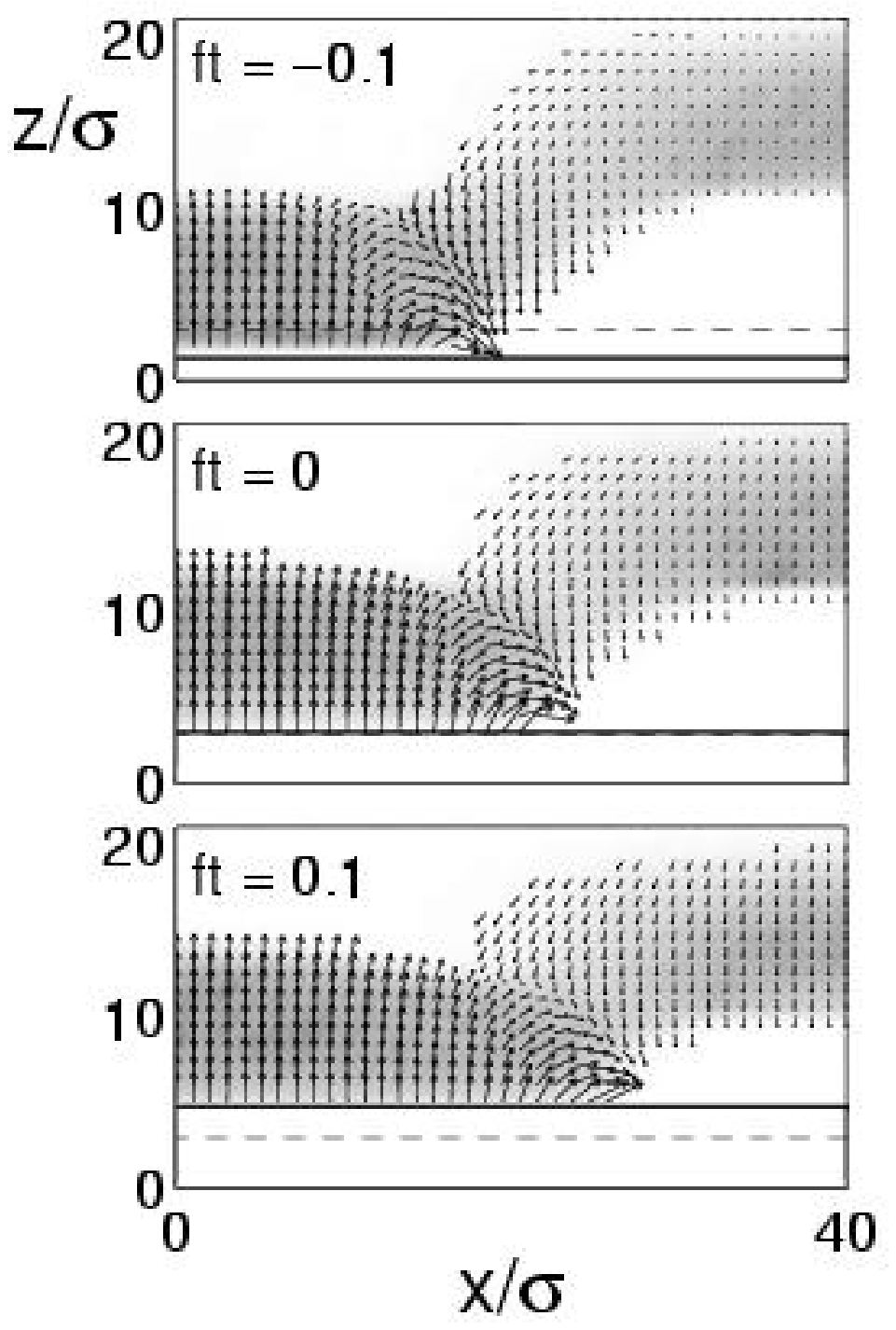}
\includegraphics[width=.493\columnwidth]{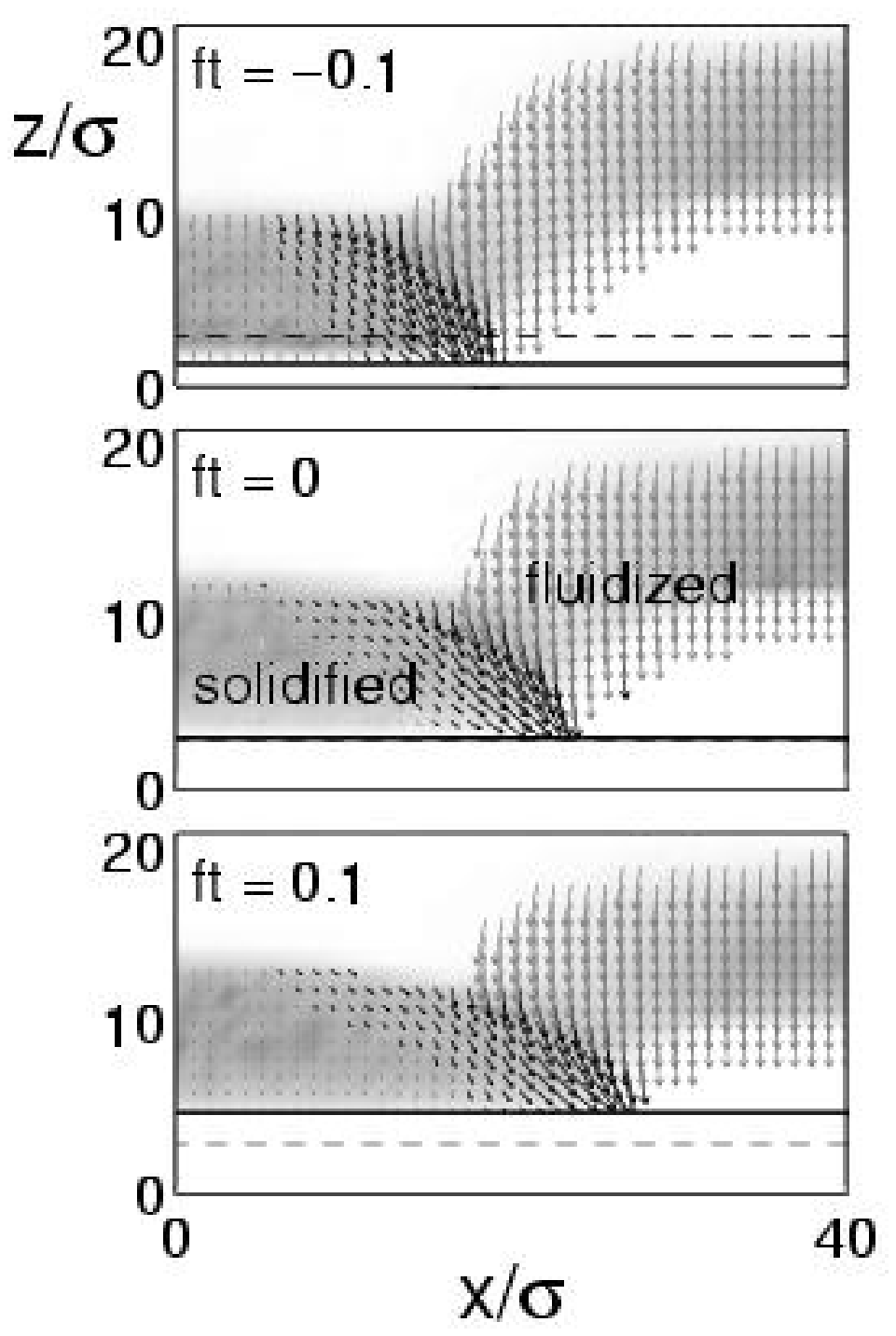}
\caption{ \label{Avalanche} Simulations reveal an avalanche occurring
inside the layer, along the interface between the solidified and
fluidized parts: (a) Instantaneous grain velocities ${\bf v}({\bf
r}(t),t)$ in the laboratory frame (indicated by arrows), averaged
over the shorter horizontal direction, when the left side of a kink
is being pushed up by the bottom plate (a horizontal solid line)
for the same case as in Fig.~\ref{ShowConvection}. A gray
background represents the layer. (b) Instantaneous velocities
in the container frame, during the same time. As the left side
is pushed up, it solidifies on the container bottom, while the
right side is fluidized and cascades down onto the left side,
as shown by gray arrows. ($a_{max} = 5.2g, f = 69$ Hz).
}
\end{figure}

As $V_{max}$ increases, qualitative changes occur in the motion of
large grains: The down-flow channel region dilates more (i.e.,
the granular volume fraction decreases) so that there is enough room for
large grains to circulate with the convecting smaller grains throughout
the volume of the trapping region, disappearing in the kink and
popping up from the upward flow region of the convection rolls.
The maximum number of trapped particles in this regime is not
determined by the area of the trapping zone;
we have not investigated this regime.

We observe in the experiment that when a kink moves too rapidly,
large grains leak out of the kink; there is a maximum speed of a
kink for which it can convey large grains. The mechanism in
Fig.~\ref{TwoSizeLayer} suggests that if a kink moves more than
the trapping zone width during a cycle, the large grains would not
follow the motion; the experimental observations are consistent in
that when the speed of a kink exceeds approximately twice the
layer depth (comparable to the zone width in Fig.~\ref{Transport})
per time $2T$, large grains are left behind the kink. 

{\it Internal avalanches and convection ---} An examination of the
instantaneous velocity field ${\bf v}({\bf r}(t),t)$ reveals that the
convection is driven by avalanches oscillating perpendicular to the
kink. When the left side of a kink is pushed up, the right side
cascades down onto the left side of the kink (Fig.~\ref{Avalanche}
(a)). These avalanches inside the layer are visualized better when the
velocity field is viewed in the container frame (Fig.~\ref{Avalanche}
(b)): When the left side is pushed up, it solidifies and moves with
the same velocity as the container until it takes off from the
container.  Meanwhile, the right side is still falling and is
fluidized. This leads to a density gradient between the solidified and
the fluidized regions. Grains flow along the surface of the solidified
region, toward the right side. We refer to this motion inside the
layer as an {\em internal} avalanche, to contrast with the usual
avalanche, which occurs on a surface~\cite{rajchenbach02}.
(Such an internal avalanche was suggested by Laroche {\it et al.}
~\cite{laroche} as a mechanism for convection in heaping.) In the
following cycle, another internal avalanche occurs in the opposite
direction, pushing grains back to the left side and to lower
heights. The oscillatory internal avalanches result in the downward
flow at the kink.  At impact with the container, grains at the bottom
of the layer move away from the kink in a horizontal direction and
later rise, forming the upward flow at the outer side of the rolls.
The trapping zone width (or convection roll size) is determined by
the horizontal displacements of such grains.

The convective motion that has been observed previously in subharmonic
standing wave patterns in oscillated layers~\cite{bizondiffusion}
can be explained by the same mechanism: when a patterned layer
collides with the container, large density gradients form along
the pattern inside the layer. Internal avalanches perpendicular to
these gradients lead to convection rolls.

{\it Conclusions ---}
We have described a form of convection in granular media that is driven
not by interaction of the grains with air or the container sidewalls
but by the intrinsic dynamics of the layer when a kink is present.
The kink and internal
avalanches that drive this convective motion are unique to oscillating
granular media because the solidification and fluidization during
each cycle that occur in the granular layer do not occur in
an oscillated liquid layer.

The kinks that lead to the avalanche motion and convection are ubiquitous
in oscillated granular layers at high enough container accelerations
($a_{max} > 7g$); hence kink-associated convection and trapping 
of large particles are generic features of bidisperse oscillated granular
layers in that regime. Since the location of a kink can be controlled by
modulation of the container oscillation frequency, it is possible to
harvest the segregated larger particles in a controlled way by sweeping
a kink to the edge of a container.

This work was supported by the Engineering Research Program of the
Office of Basic Energy Sciences of the U. S. Department of Energy
(Grant No. DE-FG03-93ER14312) and The Texas Advanced Research Program
(Grant No. ARP-055-2001).

\end{document}